\documentclass{XrU2005}
\usepackage{graphicx}

\newcommand{\mc}{\multicolumn}

\title{X-ray and soft gamma-ray behaviour of the Galactic source 1E 1743.1--2843}
\author[1]{\mbox{M. Del Santo}}
\author[2]{\mbox{L. Sidoli}}
\author[1]{\mbox{A. Bazzano}}
\author[1]{\mbox{M. Cocchi}}
\author[1]{\mbox{G. De Cesare}}
\author[3]{\mbox{A. Paizis}}
\author[1]{\mbox{P. Ubertini}}
\affil[1]{Istituto di Astrofisica Spaziale e Fisica cosmica di Roma -- INAF, via del Fosso del Cavaliere 100, 00133 Roma, Italy}
\affil[2]{Istituto di Astrofisica Spaziale e Fisica cosmica di Milano -- INAF, via E. Bassini 15, 20133 Milano, Italy}
\affil[3]{ISDC, Chemin d'$\acute{E}$cogia 16, 1290 Versoix, Switzerland}

\begin{document}
\vspace{-0.5cm}

\maketitle

\begin{abstract}
\vspace{-0.5cm}
The X-ray persistent source 1E 1743.1--2843, located in the Galactic Centre region, has been detected 
by all X-ray telescope above 2 keV, whereas it is not visible in the soft X-rays (i. e. {\it Rosat})
because of the high column density along the line-of-sight. Moreover, the nature of this source remains still unknown. 
The gamma-ray satellite {\it INTEGRAL} has long observed the Galactic Centre region in the framework of the Core Programme.
We report on results of two years of {\it INTEGRAL} observations of 1E 1743.1--2843 detected for the first time 
in the soft gamma-ray band. Since the source does not show any evidence for strong variability, we present
the broad-band spectral analysis using not simultaneous {\it XMM-Newton} observations. 
\end{abstract}
\vspace{-0.5cm}
\section{Introduction}
\vspace{-0.5cm}
1E 1743.1-2843 is one of the most absorbed ($N_{H} > 10^{23}$) X-ray sources of the Galactic Centre (GC) region, 
suggesting a distance close or even greater than the GC.
In the last years the source has been observed by numerous X-ray telescopes up to 20 keV,
but it has never been detected in the hard X-rays, because of the lack of combined high
spatial resolution and good sensitivity instruments at high energies.
{\it BeppoSAX} has long observed the Galactic Centre region, but it has never detected any bursting activity 
(in't Zand 2000), nor periodic pulsation from this source (Cremonesi et al. 1999). 
Also {\it XMM-Newton} observations reported by Porquet et al. (2003), limited to the
range below 10 keV, did not solved the mystery of the source nature;
they underlined that high energy observations could help in the
determination of the compact object nature.
We report here on the first detection in the soft gamma-ray domain (up to 70 keV) obtained during a two years 
monitoring with the gamma-ray imager IBIS, 
on-board the {\it INTEGRAL} satellite. A broad band spectral analysis 
has been also performed using re-analysed {\it XMM-Newton} data. 
\vspace{-0.5cm}
\section{Observations and Data Analysis}
\vspace{-0.5cm}
We have analysed public IBIS/ISGRI observations of the Galactic Centre region performed
in 2003 and the 2004 observations of Core Programme.
The 2003 effective exposure time is $\sim$2 Ms; $\sim$1 Ms in 2004.
Data were reduced using OSA 5.0. The 20-40 keV temporal behaviour has been extracted from the whole
data set while spectral analysis concern only the 2003 pointings.
Searching the source field in the {\it XMM-Newton} public archive 
we found 2 {\it XMM-Newton} observations performed on 2000
September $19^{th}$ (obs. 401) and  $21^{th}$ (obs. 501). Among these, only one of the {\it XMM-Newton}
observations (obs. 401) has been reported in literature (Porquet et al. 2003).
Here we present a re-analysis of this observation, now non-affected by pile-up problems.
{\it XMM-Newton} data have been analysed by SAS 6.5. In order to exclude 
pile-up effects, we selected an annular region with inner radius of 10'' and outer radius of 40''.
Background spectra were obtained from source-free regions of the same observations.

\begin{figure}[!b]
\centering
\includegraphics[width=3.5cm,height=6.5cm,angle=90]{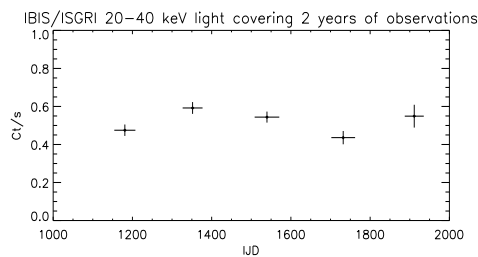}
\caption{20-40 keV temporal behaviour of 1E 1743.1--2843 with {\it INTEGRAL}starting from February 2003 till April 2005.
\label{fig:light}}
\end{figure}

\begin{table*}[!t]
\small
\caption[]{Model parameters obtained by the broad-band spectral fit using 
XMM-Newton (two observations) and ISGRI (2003 mean spectrum).
The meaning of the symbols is the following:
{\em pow}=power-law, {\em bb}=blackbody, {\em dbb}=disk-blackbody in {\sc xspec}; 
$\alpha$ is the
power-law photon index, $kT$ is the blackbody temperature
or the inner disk temperature, depending on the adopted model.
}
\begin{center}
\begin{tabular}[c]{lcccccc}
\hline\noalign{\smallskip} 
Model  &\mc{1}{c}{$N{\rm _H}$}                 &\mc{1}{c}{$kT$}   & \mc{1}{c}{$\alpha$} &\mc{1}{c}{$F^{a}_{(2-10)}$}   &\mc{1}{c}{$F^{b}_{(1-100)}$} & $\chi^2$/dof \\

       &($10^{22}$~cm$^{-2}$)  &              \mc{1}{c}{(keV)}                       & &                &               & \\
\noalign{\smallskip\hrule\smallskip}
Obs. 401    & & &  & &    \\
bb+po               & $19.5 ^{+1.1}_{-0.9}$  &  $1.8^{+0.1}_{-0.1}$& $3.1^{+0.1}_{-0.1}$ &  3.9    & 7.3 &  468/375  \\ 
diskbb+po           & $18.6 ^{+0.8}_{-0.7}$ & $3.1^{+0.1}_{-0.1}$  & $2.5^{+0.1}_{-0.2}$  &  3.5    &      &  475/375  \\
\noalign{\smallskip\hrule\smallskip}
Obs. 501    & & &  & &    \\
bb+po               & $18.0^{+1.3}_{-1.0}$  & $1.6^{+0.1}_{-0.1}$ &  $3.3^{+0.1}_{-0.1}$  & 3.2 & 5.9  & 392/323  \\
diskbb+po           & $19.8^{+0.9}_{-0.9}$ & $2.6^{+0.2}_{-0.2}$ &  $2.9^{+0.1}_{-0.1}$  & 3.3  &    & 401/323  \\

\noalign{\smallskip\hrule\smallskip}
\end{tabular}
\small
\begin{itemize}
\small
\item[$a$]The 2--10 keV flux of the unabsorbed fit model in units of 10$^{-10}$~erg~cm$^{-2}$~s$^{-1}$.
\vspace{-0.2cm}
\small
\item[$b$]The broad-band flux (1--100 keV) of the unabsorbed best-fit model in units of 10$^{-10}$~erg~cm$^{-2}$~s$^{-1}$.
\end{itemize}
\vspace{-0.8cm}
\label{tab:fit}
\end{center}
\end{table*}
\vspace{-0.5cm}

\section{Results and discussion}
\begin{figure}[!b]
\centering
\includegraphics[width=4.5cm,height=6.5cm,angle=-90]{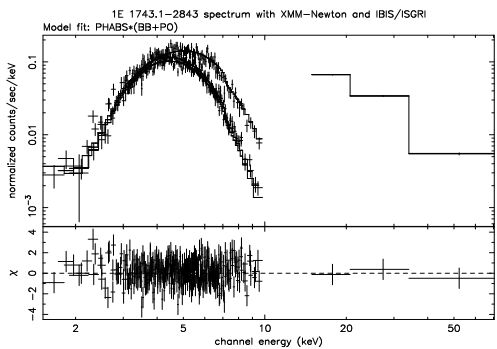}
\caption{Count rate absorbed spectrum, BB+PO model fit and residuals of MOS1, MOS2, PN (obs. 501) and ISGRI data.
\label{fig:spec}}
\end{figure}

\vspace{-0.5cm}
The Galactic Centre is pointed by {\it INTEGRAL} during two visibility windows per year.
The temporal behaviour of 1E 1743.1--2843 over 2 years of IBIS observations for a total 
of about 3 Ms is shown (Fig. \ref{fig:light}). The source shows marginal variability over few months 
times scale, in agreement with results reported by Belanger et al. (2005).  
Because of its rather constant high energy behaviour, we fitted the average IBIS/ISGRI spectrum of 2003
with non-simultaneous PN, MOS1 and MOS2 data (Fig. \ref{fig:spec}). We used two models: a black body plus
a power law and a multi-temperature disc plus a power law. 
Spectral parameters are presented in Tab. \ref{tab:fit}.   
The two observations show parameters consistent within the errors. 
The steepness of the power law component indicates a soft hard-X-ray spectrum for 
this source. We confirm the further indication by Cremonesi et al. (1999)
which rules out the HMXB nature.
So far, the observational scenario seems to be in favour of a LMXB system.
Starting from this hypothesis, the nature of the compact object needs to be discussed.
We have estimated luminosities 
and related fractions in Eddington luminosity both for Neutron Star (NS) and Black Hole (BH),
considering three possible distances for 1E 1743.1--2843.
We assumed M$_{NS}$ = 1.4 M$_{\odot}$ and M$_{BH}$ = 10 M$_{\odot}$.\\
{\bf Let's suppose that the accreting object is a NS}.\\
During more than 20 years of observations, the lack of type-I X-ray bursts is noteworthy.
Nevertheless it is in agreement with the estimated luminosities in Eddington luminosity fractions (Tab. \ref{tab:lumi}).
Type-I X-ray bursts become rare going up a few percent of Eddington luminosity (Lewin et al. 1995).
So, in this first case we have 2 possibilities: the NS is a rare burster from which 
we did not detect any thermonuclear flash or the system is a bright LMXRB located
behind the Galactic Centre at a distance at least $>$15 kpc.\\
{\bf Let's suppose that the accreting object is a BH}.\\ 
By our spectral parameters, 1E 1743.1--2843 should be  
a BH binary in the canonical high/soft state, contrary to the low/hard state 
proposed by Porquet et al. (2003).
BH binaries in the soft state show luminosities as at least a few percent 
of L$_{Edd}$ (Maccarone 2003).
In this case the source distance cannot be less than 20 kpc.
Considering that nearly all LMXRBs with persistent X-ray emission contain
a NS (van der Klis 2004) and the strong variability usually associated to BH binary systems,
this assumption seems to be less strong than the NS one.\\
Persistency, BB temperatures and faint and steep power law component
support the NS nature for this source.
\vspace{-0.5cm}
\begin{table}[!t]
\caption[]{Luminosities of 1E 1743.1--2843 at different distances calculated both for a NS and BH.}
\begin{center}
\begin{scriptsize}
\begin{tabular}[c]{lccc}
\hline\noalign{\smallskip} 
\mc{1}{c}{Distance} &\mc{1}{c}{L$_{1-100 keV}$} &\mc{1}{c}{L/L$_{Edd}$ (BH)}  &\mc{1}{c}{L/L$_{Edd}$ (NS)}\\
 & (erg/s) & & \\
\noalign{\smallskip\hrule\smallskip}
8.5 kpc & 5.2 $\times$ 10$^{36}$ & 0.3\% & 3\%\\ 
\noalign{\smallskip\hrule\smallskip}
12 kpc  & 1.0 $\times$ 10$^{37}$ & 0.8\% & 5\%\\
\noalign{\smallskip\hrule\smallskip}
20 kpc     &  2.9 $\times$ 10$^{37}$ & 2.0\%& 11\% \\
\noalign{\smallskip\hrule\smallskip}
\end{tabular}
\end{scriptsize}
\small
\label{tab:lumi}
\end{center}
\vspace{-0.8cm}
\end{table}

\end{document}